\documentclass{article}
\usepackage{spconf,amsmath,graphicx}

\usepackage{cite}
\usepackage[utf8]{inputenc} 
\usepackage[T1]{fontenc}    
\usepackage{hyperref}       
\usepackage{url}            
\usepackage{booktabs}       
\usepackage{amsfonts}       
\usepackage{nicefrac}       
\usepackage{microtype}      
\usepackage[dvipsnames]{xcolor} 
\usepackage{adjustbox}
\usepackage{multirow}
\usepackage{multicol}
\usepackage{tabularx}

\title{Leave No Knowledge Behind During Knowledge Distillation: \\ Towards Practical and Effective Knowledge Distillation for Code-Switching ASR Using Realistic Data}
\name{Liang-Hsuan Tseng\(^\dagger\), Zih-Ching Chen\(^\star\), Wei-Shun Chang\(^\dagger\),}
\namenxt{Cheng-Kuang Lee\(^\star\), Tsung-Ren Huang\(^\dagger\), Hung-yi Lee\(^\dagger\)}
\address{\(^\dagger\)National Taiwan University, \(^\star\)NVIDIA AI Technology Center, NVIDIA, Taipei, Taiwan }
\begin{document}
%
\maketitle
\begin{abstract}

Recent advances in automatic speech recognition (ASR) often rely on large speech foundation models for generating high-quality transcriptions. 
However, these models can be impractical due to limited computing resources. 
The situation is even more severe in terms of more realistic or difficult scenarios, such as code-switching ASR (CS-ASR). 
To address this, we present a framework for developing more efficient models for CS-ASR through knowledge distillation using realistic speech-only data. 
Our proposed method, Leave No \textbf{K}nowledge Behind During \textbf{K}nowledge \textbf{D}istillation (K\(^2\)D), leverages both the teacher model's knowledge and additional insights from a small auxiliary model. 
We evaluate our approach on two in-domain and two out-domain datasets, demonstrating that K\(^2\)D is effective. By conducting K\(^2\)D on the unlabeled realistic data, we have successfully obtained a 2-time smaller model with 5-time faster generation speed while outperforming the baseline methods and the teacher model on all the testing sets. 
We have made our model publicly available on \href{https://huggingface.co/andybi7676/k2d-whisper.zh-en}{Hugging Face}. 
\end{abstract}


\section{Introduction}
\label{sec:intro}

ASR has long posed significant challenges within the speech community.
Recent breakthroughs have leveraged techniques such as self-supervised learning (SSL)~\cite{wav2vec2, hubert, bestrq}, self-training~\cite{self_training_meta, iterative_pseudo_labeling, simIPL}, and large-scale or weakly-supervised learning~\cite{whisper, google_usm, seamlessm4t} to develop high-quality ASR systems.
Despite these advancements, achieving high-quality transcriptions often requires the use of large speech foundation models~\cite{whisper, google_usm, seamlessm4t}.
This can be impractical for many applications, especially when computational resources are limited.
The challenge is further amplified in realistic scenarios, where speech is more diverse and difficult to transcribe.
A particularly challenging example of this diversity is code-switching ASR (CS-ASR), where speakers frequently alternate between languages within and between utterances.

To address this issue, we aim to develop smaller and faster models for CS-ASR using realistic data that we have collected. Our dataset comprises academic course videos covering a range of subjects, including but not limited to engineering, science, and liberal arts.
The data comprises approximately 60,000 hours of raw speech, mostly Mandarin-English code-switched.
Despite the enormous volume of audio data, we lack transcriptions for training, which presents a significant challenge in developing effective ASR models. 
To address this, we propose using pseudo-labeling as a solution to leverage the unlabeled data. 
With the pseudo labels, we can conduct knowledge distillation along with the teacher model~\cite{distil-whisper}. 
However, pseudo-labels often contain errors and hallucinations~\cite{self_training_meta}, which can degrade model performance if not properly managed.

\begin{figure}[hb!]
    \centering
    \includegraphics[width=0.95\columnwidth]{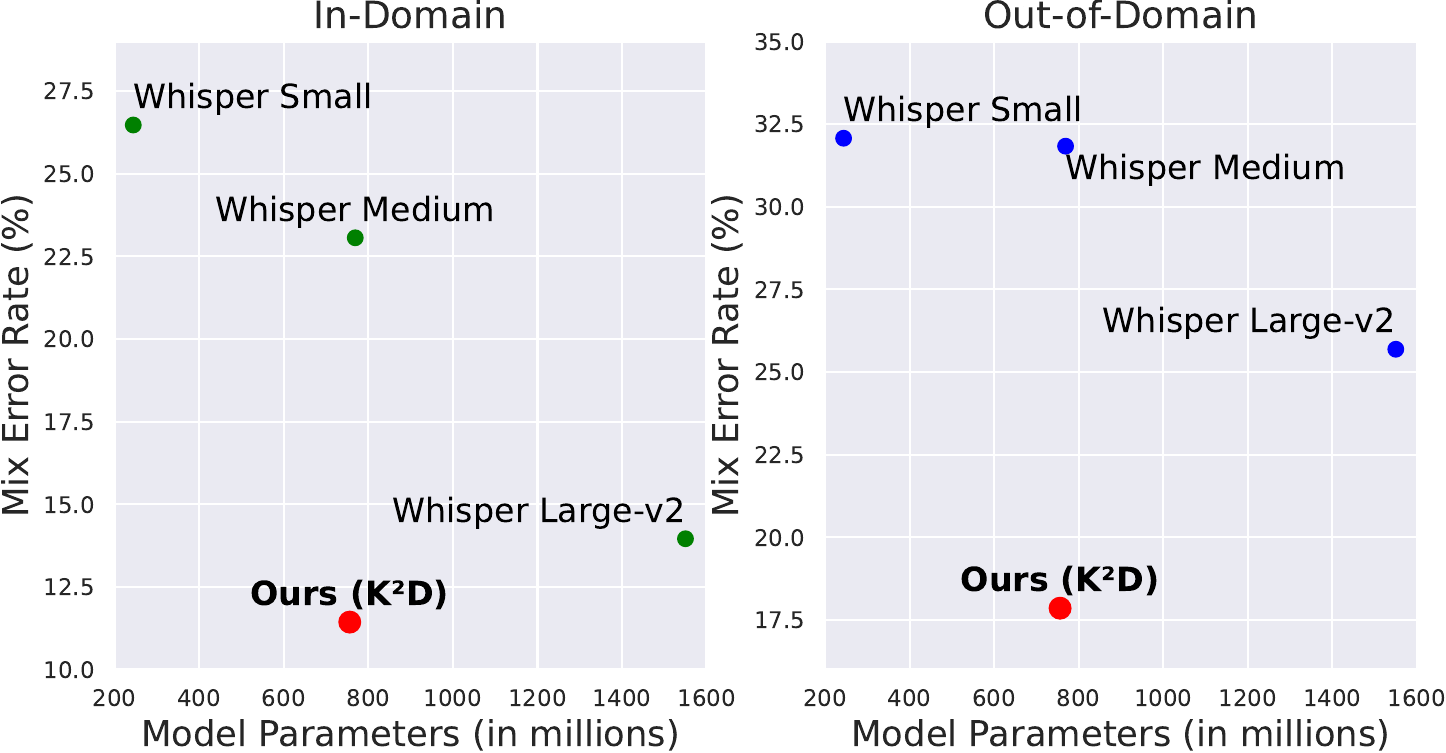}
    \caption{
        Our proposed framework K\(^2\)D achieves significant performance improvement over the teacher model (Whisper Large-v2) on both in-domain and OOD testing sets. 
    }
    \label{fig:intro_demo}
\end{figure}

This work proposes a novel method called Leave No \textbf{K}nowledge Behind During \textbf{K}nowledge \textbf{D}istillation (K\(^2\)D). 
Our proposed method enhances the traditional knowledge distillation process by integrating insights from both a large teacher model and a smaller auxiliary model. 
We filter data by referencing the small model's transcriptions on the realistic data. 
By evaluating on two in-domain and two out-of-domain (OOD) testing sets, we provide evidence that applying K\(^2\)D can lead to a better student model than the original knowledge distillation. 
The distilled models even surpass the teacher model on all the testing sets while being two times smaller and five times faster, showing the strong generalizability and efficiency of applying K\(^2\)D (Fig.~\ref{fig:intro_demo}). 
To our best knowledge, we are the first to explore knowledge distillation for CS-ASR with unlabeled realistic data, and we propose an improved framework by incorporating a small auxiliary model. 
\vspace{-2pt}

\section{Background}
\label{sec:background}
\subsection{Code-Switching ASR}
\label{subsec:code-switching_ASR}
Code-switching ASR (CS-ASR) is a challenging sub-problem in automatic speech recognition. 
Unlike monolingual utterances, a code-switched utterance may contain transitions between languages at one or multiple positions.
The problem is particularly pronounced in regions with more than one official language. 
Conventional methods address this issue by incorporating language identification (LID)~\cite{ctc_lid2019, lid_02, lid_03, lid_lhz, lid_05}, using monolingual data~\cite{cs_mono01, cs_mono02, cs_mono03, cs_mono04, cs_mono_lab}, or leveraging data augmentation techniques~\cite{cs_data_aug_lab, cs_data_aug_02}. 
Recent approaches focus on utilizing large language models (LLMs) to generate CS data or investigating large speech foundation models CS-ASR with different techniques~\cite{cs_llm_01, cs_llm_lab, cs_llm_03}. 
Our work differs from those mentioned above by optimally utilizing large-scale unlabeled CS data collected from real-world scenarios.

\subsection{Knowledge Distillation}
\label{subsec:bg_knowledge_distillation}
Knowledge distillation (KD)~\cite{kd_hinton} has been a common and effective method for model compression. 
In KD, a student model learns to mimic the behavior of a stronger teacher model to achieve similar performance.
With the advent of powerful large foundation models, KD has been employed across multiple fields to obtain smaller and more efficient models~\cite{distilbert, distilhubert, distilvlm, efficientvlm}, further strengthening its effectiveness.
Typically, the targets of KD can be categorized into two types: the teacher's output and hidden representations. Distilling the teacher's hidden representation is common when the task is not specific~\cite{distilbert, distilhubert, distilvlm}. However, these methods are not tailored for CS-ASR and do not leverage large-scale unbalanced data effectively. In this work, we address these limitations by focusing specifically on CS-ASR in real-world scenarios and employing the learning objectives of Distil-Whisper~\cite{distil-whisper}, which utilize the teacher model's outputs during knowledge distillation. Our approach optimally uses large-scale unlabeled CS data, enhancing the efficiency and performance of the distilled models for CS-ASR.

\subsection{Large-Scale Pseudo-Labels for Distilled Models}
\label{subsec:distil-whisper}
Inspired by the success of Whisper~\cite{whisper}, Distil-Whisper~\cite{distil-whisper} aims to provide a more efficient model through knowledge distillation using large-scale pseudo-labels. 
Distil-Whisper utilizes 21,170 hours of audio data from 9 publicly available datasets for KD. 
However, the datasets are monolingual and exclusively in English, which motivates us to develop our own distilled model for more effective generations in a code-switching context. 
Unlike Distil-Whisper, which utilizes data in labeled monolingual corpora for training, the data we collected is code-switched and has no true labels. 
Therefore, we cannot filter data based on the word error rate (WER) between the real and the pseudo-labeled transcriptions, as suggested in Distil-Whisper. 
To address the issue, we aim to provide a simple and effective method based on a small auxiliary model to perform cross-model validation for data filtering without any labeled data.

\subsection{Semi-Supervised Learning and Self-Training}
\label{subsec:semi-supervised_learning_and_self-training}
Semi-supervised learning and self-training focus on developing techniques to exploit the unlabeled data. 
In semi-supervised learning, the data can be separated into labeled and unlabeled ones.
Utilizing the labeled data for supervised learning is trivial. 
Thus, most efforts are dedicated to leveraging the unlabeled data, which is often large-scale and noisy. 
One of the most common solutions is to use the supervised-learned model to generate pseudo-labels and then use the pseudo-labels for self-training~\cite{self-training}. 
Due to the model's imperfections, directly leveraging all pseudo-labels may lead to slight improvement. 
To address the issue, previous works may use additional LM to improve pseudo-labels quality~\cite{self_training_meta} or conduct data filtering~\cite{st_data_filtering_01, st_data_filtering_02, st_data_filtering_03, st_data_filtering_04}. 
Moreover, recent works propose techniques like model ensembling, iterative pseudo-labeling, or continuous pseudo-labeling for self-training~\cite{self_training_meta, simIPL, momentum_pl, continuous_pl}. 
These methods are more effective but often involve complicated algorithms during training, resulting in even higher computation resources. 

\section{Method: K\(^2\)D}
\label{sec:method_K2D}
\begin{figure*}[htb!]
    \centering
    \includegraphics[width=2\columnwidth]{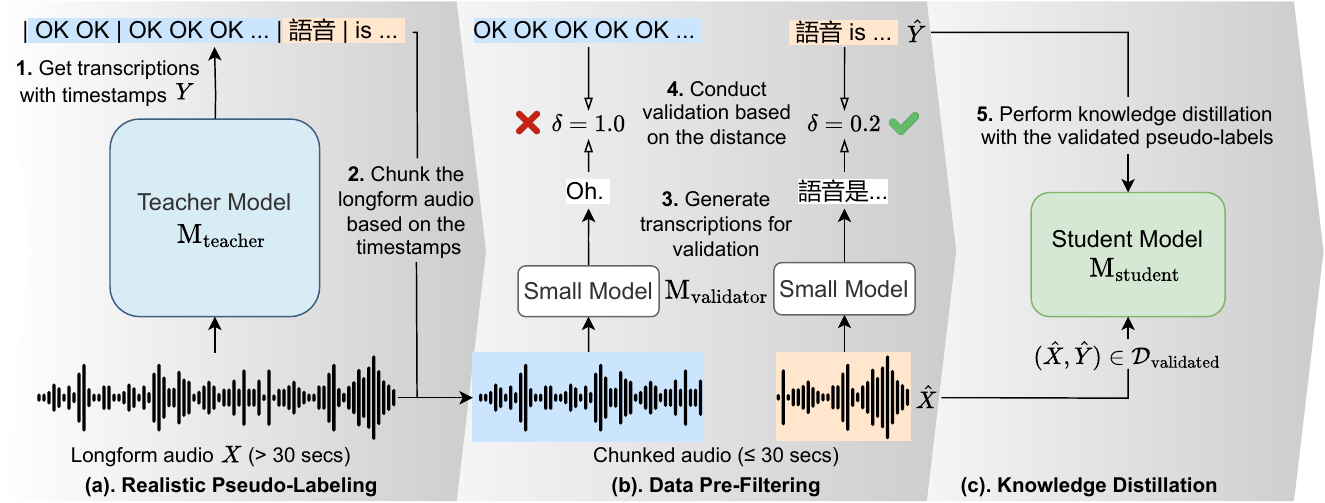}
    \caption{
    Overview of the K\(^2\)D Framework. (a) \textbf{Realistic Pseudo-Labeling:} The teacher model generates transcriptions with timestamps from long-form audio. (b) \textbf{Data Pre-Filtering:} Chunked audio is validated by the small auxiliary model, filtering out inaccurate labels. (c) \textbf{Knowledge Distillation:} Validated pseudo-labels are used to train the student model, enhancing accuracy and efficiency.
    }
    \label{fig:framework}
\end{figure*}
We propose a simple and effective framework called K\(^2\)D for developing an efficient code-switching ASR system via knowledge distillation using speech-only realistic data. 
Our K\(^2\)D framework comprises three stages: realistic pseudo-labeling, data pre-filtering, and knowledge distillation. First, we perform \textbf{realistic pseudo-labeling} (\S~\ref{subsec:realistic_pseudo_labeling}) on the realistic long-form data and segment each into small chunks. 
Next, we conduct \textbf{data pre-filtering} (\S~\ref{subsec:data_pre-filtering}) to validate the chunkwise pseudo-labels based on the additional knowledge from the auxiliary small model.
Finally, we use the validated data for \textbf{knowledge distillation} (\S~\ref{subsec:knowledge_distillation}). 
The illustration of the proposed framework is presented in Fig.~\ref{fig:framework}.

\subsection{Realistic Pseudo-Labeling}
\label{subsec:realistic_pseudo_labeling}

Given a long-form realistic speech \(X\) and the teacher model \({\rm M}_{\rm teacher}\), we first generate the corresponding transcriptions with timestamps \(Y={\rm M}_{\rm teacher}(X)\). 
Specifically, we use Whisper~\cite{whisper} as the teacher model, which produces timestamps along with the transcriptions when performing sequential generation on the long-form audio. 
With an audio-transcription pair \((X, Y)\), we then generate \(M\) segments based on the timestamps in \(Y\). 
The segments can be formulated as \([(X'_1, Y'_1, d'_1), (X'_2, Y'_2, d'_2), \ldots, (X'_M, Y'_M, d'_M)]\), where each segment tuple \((X'_i, Y'_i, d'_i), \forall i \in [1, M]\) represents the audio, transcription, and the duration of the \(i\)-th segment. 

Next, we concatenate the adjacent segments into chunks with a maximum duration of 30 seconds based on \([d'_1, \ldots, d'_M]\). 
We iterate through each segment in order, add the segment to the current chunk, and update the chunk's duration, provided that doing so does not make the current chunk exceed 30 seconds. 
If adding a segment increases the duration of the chunk to more than 30 seconds, we finalize the current chunk and start a new chunk with the current segment.
This process continues until all segments have been processed, ensuring that each chunk has a duration \(\leq\) 30 seconds. 
We describe the chunked audio-transcription pairs as \([(\hat{X}_1, \hat{Y}_1), \ldots, (\hat{X}_N, \hat{Y}_N)], N \leq M\). 

Finally, let \(f\) be a function that generates the set of chunked audio-transcription pairs given \(X\); we can construct the realistic pseudo-labeled dataset based on the original dataset \(\mathcal{X}\):
\[
    \mathcal{D}_{\rm RPL} = \bigcup_{\forall X \in \mathcal{X}} f(X), 
    \quad 
    f(X)=\{(\hat{X}_i, \hat{Y}_i) \mid i \in [1, N] \}. 
\]
We keep all the timestamp tokens in the transcription \(\hat{Y}\) to facilitate segmentation behavior learning during knowledge distillation. 





\subsection{Data Pre-Filtering}
\label{subsec:data_pre-filtering}
Due to the imperfection of the teacher model, the pseudo-labeled audio-transcription pairs \((\hat{X}, \hat{Y}) \in \mathcal{D}_{\rm RPL}\) may inevitably contain different levels of errors or hallucinations.
As mentioned in the previous works~\cite{self_training_meta, distil-whisper}, one of the most common types of hallucination is looping or repetitive generation. 
The hallucination can be detected trivially by counting the n-gram according to~\cite{self_training_meta}.
Specifically, we may consider a pseudo-label \(\hat{Y}\) as hallucinated if it contains a \(n\)-gram pattern for over \(c\) times.
We refer this detection as \textbf{\textit{trivial}} or \textbf{\textit{\(n\)-gram}} method, denoted as follows:
\[
\label{eq:ngram_detection}
    h^n_c(\hat{Y}) = 
    \begin{cases}
        1, \quad & \text{if } \hat{Y} \text{ contains a \(n\)-gram pattern over \(c\) times}, \\
        0, \quad & \text{otherwise}.
    \end{cases}
\]
By filtering out the pseudo-labels with the trivial hallucinations, we can form a new dataset:
\begin{align}
\label{eq:D_trivial}
    \mathcal{D}_{\rm trivial} = 
    \{ (\hat{X}, \hat{Y}) \in \mathcal{D}_{\rm RPL} \mid h^n_c(\hat{Y}) = 0 \}.
\end{align}
While being suitable for detecting the looping errors, the \(n\)-gram method might fall short for the other types of hallucinations.

In this work, we aim to provide a simple and effective method to validate the pseudo-labels by introducing a small auxiliary model \({\rm M}_{\rm validator}\) for validation. 
Specifically, we perform data filtering by validating across the outputs from the teacher model \({\rm M}_{\rm teacher}\) and the small auxiliary model \({\rm M}_{\rm validator}\). 
Given a pseudo-labeled audio-transcription pair \((\hat{X}, \hat{Y})\), we first generate the validation-oriented transcription \(\hat{V}={\rm M}_{\rm validator}(\hat{X})\) with the small model, and then perform cross-model validation between \(\hat{Y}\) and \(\hat{V}\). 
The cross-model validation is accomplished by calculating the distance metric \(\delta\), accompanied by an additional threshold \(\alpha \in [0.0, 1.0)\) for the final determination. 
We formulate the filtered dataset after cross-model validation as follows:
\begin{align}
\label{eq:D_validated}
    \mathcal{D}_{\rm validated} = \{ (\hat{X}, \hat{Y}) \in \mathcal{D}_{\rm RPL} \mid \delta(\hat{Y}, \hat{V}) \leq \alpha \}.
\end{align}

As presented, the distance metric \(\delta\) is essential during filtering. 
We introduce the three different kinds of distance metrics we experiment with. 
First, we introduce the two direct measurements. 
We define the distance metric by directly calculating the mixed error rate (MER) between the teacher's transcription \(\hat{Y}\) and the validator's transcription \(\hat{V}\), or the phoneme error rate (PER) between the phonemicized transcriptions \(\hat{Y}_{\rm p}\) and \(\hat{V}_{\rm p}\): 
\begin{table}[h!]
\centering
\begin{minipage}[b]{0.49\linewidth}
    \begin{align}
        \label{eq:delta_mer}
        \delta_{{\rm MER}} &= {\rm MER}(\hat{Y}, \hat{V}),
    \end{align}
\end{minipage}
\hfill
\begin{minipage}[b]{0.49\linewidth}
\begin{align}
    \label{eq:delta_per}
    \delta_{{\rm PER}} &= {\rm PER}(\hat{Y}_{\rm p}, \hat{V}_{\rm p}),
\end{align}
\end{minipage}
\hfill
\end{table}

Last but not least, we introduce the composite metric, which explicitly considers the trivial hallucinations of \(\hat{Y}\) and \(\hat{V}\):
\begin{align}
\label{eq:delta_composite}
    \delta_{\rm comp}=
    \max( h^n_c(\hat{Y}),
    \min(1 - h^n_c(\hat{V}), \delta_{\rm PER}(\hat{Y}_{\rm p}, \hat{V}_{\rm p}) ) ). 
\end{align}
When \(\hat{Y}\) is trivially hallucinated, we should directly discard the pseudo-labels; on the other hand, we should keep the pseudo-labels when \(\hat{V}\) is trivially hallucinated. 


\subsection{Knowledge Distillation}
\label{subsec:knowledge_distillation}
In K\(^2\)D, we conduct knowledge distillation on the cross-validated realistic dataset \(\mathcal{D}_{\rm validated}\) (Eq.~\ref{eq:D_validated}). 
Otherwise, the distillation process mainly follows Distil-Whisper~\cite{distil-whisper}. 
First, we construct the student model \({\rm M}_{\rm student}\). 
The student model is an encoder-decoder model, in which we initialized parameters from the teacher model \({\rm M}_{\rm teacher}\). 
Given a pseudo-labeled audio-transcription pair \((\hat{X}, \hat{Y}) \in \mathcal{D}_{\rm validated}\), we can generate student's prediction \(\hat{S} = [\hat{s}_1, \hat{s}_2, \ldots, \hat{s}_k]\) based on \({\rm M}_{\rm student}\) and teacher forcing with \(\hat{Y}\);
and calculate the cross-entropy loss as follows:
\[
\mathcal{L}_{\rm CE} = - \sum_{i=1}^{k} \log p(\hat{s}_i = \hat{y}_i \mid \hat{y}_{<i}, \hat{X}),
\]
where \(p\) indicates the probability function. 
Moreover, we generate the prediction distribution \(\hat{Q} = {\rm M}_{\rm teacher}(\hat{X}, \hat{Y})\). 
The distribution \(\hat{Q}=\hat{q}_{1:k}\) then served as the soft-label during knowledge distillation. 
We use the soft-label \(\hat{Q}\) from the teacher model and the probability distribution \(\hat{P} = p(\hat{S}) = \hat{p}_{1:k}\) to calculate the Kullback-Leibler (KL) divergence loss between \(\hat{Q}\) and \(\hat{P}\): 
\[
\mathcal{L}_{\rm KL} = \sum_{i=1}^{k} {\rm KL}(\hat{q}_i, \hat{p}_i).
\]

Finally, we combine the cross-entropy loss and the KL divergence loss. 
The overall objective is defined as follows:
\begin{align}
\label{eq:kd_objective}
    \mathcal{L}_{\rm KD} = \beta \mathcal{L}_{\rm CE} + \gamma \mathcal{L}_{\rm KL},
\end{align}
where \(\beta\) and \(\gamma\) are the weighted coefficients. 

\begin{table}[!htb]
    \centering
    \small
    \caption{The duration distribution over the top six subjects with the longest durations. Eng: Engineering; LA: Liberal Arts; Sci: Science; EECS: Electrical Engineering and Computer Science; Mgmt: Management; BA: Bio-resource and Agriculture. 
    }
    \setlength\tabcolsep{3pt} 
    \begin{adjustbox}{width=\columnwidth*1,center}
    \begin{tabular}{l | ccccccc}
    \toprule
         Subject & Eng & LA & Sci & EECS & Mgmt & BA & Others  \\  
         \midrule
         Duration (hr) & 9540 & 7900 & 6222 & 4796 & 3374 & 3236 & 25365  \\
         \bottomrule
    \end{tabular}
    \label{tab:training_data_distribution}
    \end{adjustbox}
\end{table}

\section{Experiments}

\subsection{Data}
\label{subsec:data}
As mentioned in Section~\ref{sec:intro}, we use self-collected data for training. 
The data is collected from the course recordings of National Taiwan University, resulting in about 60,000 hours of audio. 
The courses can be categorized into different subjects, and we present the time distribution of the top six subjects with the most extended durations in Table~\ref{tab:training_data_distribution}.
As for the testing sets, we introduce them as follows:

\textbf{COOLTEST} represents our own \textbf{\textit{in-domain}} and in-house testing set. 
The testing set contains 5 hours of speech from different subjects. 
We self-annotated the testing data by revising the transcriptions generated by Whisper Large-v2. 

\textbf{NTUML2021} is a speech dataset from National Taiwan University's 2021 "Machine Learning" course. 
Since the dataset is derived from lecture videos, we use the dataset as a publicly available \textbf{\textit{in-domain}} Mandarin-English CS-ASR corpus. 
We use the 9-hour testing split to perform the evaluation. 

\textbf{CommonVoice} is a publicly available, multilingual dataset of voice recordings collected by Mozilla for ASR~\cite{commonvoice}. 
It includes contributions from volunteers worldwide, providing a diverse range of accents, languages, and demographic variations. 
The diversity of the corpora makes it suitable for serving as the \textbf{\textit{out-of-domain}} evaluation set. 
We evaluate our method on the \(16\)-th version of CommonVoice with the zh-TW split flag, indicating the speech collected from Taiwan. 

\textbf{ASCEND}~\cite{ascend} is a publicly available, spontaneous, Mandarin-English CS dataset collected from conversational recordings. 
Furthermore, it is collected in Hong Kong, which differs from the chosen CommonVoice testing split mentioned earlier. 
The dataset, which is code-switched and realistic, is suitable for serving as \textbf{\textit{out-of-domain}} evaluation of our method.

\begin{table*}[!htb]
    \centering
    \caption{The evaluation results of K\(^2\)D on the two in-domain and two out-of-domain (OOD) testing sets. Our method demonstrates clear performance improvements and strong generalizability compared with the teacher model and the baseline methods. MERR indicates the mix error reduction rate, calculated relative to the teacher model. The speed-up is calculated based on the RTF averaged over five runs. 
    }
    \renewcommand{\arraystretch}{1.0}
    \label{tab:main}
    \begin{adjustbox}{width=\columnwidth*2,center}
    \hspace*{1em}
    \begin{tabular}{lcc cc cc}
    \toprule
    \multirow{2}{*}[-3pt]{\textbf{Method}} & \multirow{2}{*}[-3pt]{\begin{tabular}{c}\textbf{Speed-}\\\textbf{up}\end{tabular}} & \multirow{2}{*}[-3pt]{\begin{tabular}{c}\textbf{Data}\\\textbf{usage}\end{tabular}} & \multicolumn{2}{c}{\textbf{In-domain MER\% (MERR) \(\downarrow\)}} & \multicolumn{2}{c}{\textbf{OOD MER\% (MERR) \(\downarrow\)}} \\ 
    \cmidrule(lr){4-5} \cmidrule(lr){6-7}
     &  &  & \textbf{COOLTEST} & \textbf{NTUML2021} & \textbf{CV16 (zhTW)} & \textbf{ASCEND} \\ 
    \midrule
    \multicolumn{3}{l}{\textbf{\textsc{Teacher Model}}} \\
    \quad Whisper Large-v2 & \(\times1.0\) & - & 13.96 \hfill (-0.0\%) & 7.35 \hfill (-0.0\%) & 9.07 \hfill (-0.0\%) & 25.69 \hfill (-0.0\%) \\
    \midrule
    \multicolumn{3}{l}{\textbf{\textsc{Baseline (KD)}}} \\
    \quad Full Data & \(\times5.0\) & 100\% & 12.98 \hfill (-7.0\%) & 6.29 (-14.4\%) & 7.80 \hfill (-14.0\%) & 22.07 \hfill (-14.1\%) \\
    \quad Trivial Method (Eq.~\ref{eq:D_trivial}) & \(\times4.9\) & 97\% & 12.56 \hfill (-10.0\%) & 6.28 (-14.5\%) & 8.05 \hfill (-11.2\%) & 19.78 \hfill (-23.0\%) \\
    \cmidrule[0.5pt](lr){1-7}
    \multicolumn{3}{l}{\textbf{\textsc{Our Method (K\(^2\)D)}}} \\
    \quad Direct \({\rm MER}\) (Eq.~\ref{eq:delta_mer}) & \(\times5.1\) & 55\% & 11.96 \hfill (-14.3\%) & 6.24 (-15.1\%) & 7.54 \hfill (-16.9\%) & 19.40 \hfill (-24.5\%) \\
    \quad Direct \({\rm PER}\) (Eq.~\ref{eq:delta_per}) & \(\times5.0\) & 61\% & 11.54 \hfill (-17.3\%) & 6.17 (-16.0\%) & \textbf{7.33} \hfill \textbf{(-19.2\%)} & 18.82 \hfill (-26.7\%) \\
    \quad Composite \(\delta_{\rm comp}\) (Eq.~\ref{eq:delta_composite}) & \(\times5.1\) & 74\% & \textbf{11.44} \hfill \textbf{(-18.1\%)} & \textbf{6.09 (-17.1\%)} & 7.62 \hfill (-16.0\%) & \textbf{17.86} \hfill \textbf{(-30.5\%)}  \\
    
    \bottomrule
    \end{tabular}
    \hspace*{1em}
    \end{adjustbox}
\end{table*}

\subsection{Models and Training Details}
\label{subsec:models}
In K\(^2\)D, we have three kinds of models: the teacher model \({\rm M}_{\rm teacher}\), the small validation model \({\rm M}_{\rm validator}\), and the student model \({\rm M}_{\rm student}\). 
Practically, we use Whisper Large-v2 as \({\rm M}_{\rm teacher}\), and Whisper Base as \({\rm M}_{\rm validator}\). 
The validator model is over 20 times smaller than the teacher model, facilitating the fast validation-oriented transcription generation on large-scale realistic data. 
On the other hand, the student model comprises 32 encoder layers and two decoder layers and is two times smaller than the teacher model. 
The parameter initialization follows Distil-Whisper, where the encoder is identical to the teacher model's; and the 2-layer decoder is initialized from the first and the last layer of the teacher's decoder. 
We perform knowledge distillation on 4 NVIDIA H100. 
We use batch size 256 and update for 120,000 steps, which takes about 42 hours of training. 
The encoder is frozen during training. 
We set the threshold \(\alpha=0.4\) in Eq.~\ref{eq:D_validated} for cross-model validation; 
while the weighted coefficients \(\beta\) and \(\gamma\) of \(\mathcal{L}_{\rm KD}\) in Eq.~\ref{eq:kd_objective} are set to \(0.8\) and \(1.0\). 
We use g2p~\cite{g2pE2019} for English and pinyin for Mandarin phonemicization.

\subsection{Evaluation Metrics}
\label{subsec:evaluation_metrics}
We use the mixed error rate (MER) to evaluate the quality of code-switching ASR. 
The metric gathers the two common evaluation metrics in the two languages: the character error rate (CER) in Mandarin and the word error rate (WER) in English. 
We perform the long-form evaluation if not specified to simulate the realistic scenario. 
Moreover, we use the real-time factor (RTF) under the same computing resource to measure the speed-up brought by the student model.


\section{Results}
\label{sec:results}

\begin{table}[!ht]
    \small
    \renewcommand{\arraystretch}{1.0}
    \caption{
        The detailed performance analysis of the two languages on the in- and out-of-domain code-switching datasets. Man. stands for the CER of the Mandarin part, while Eng. is the WER of the English part in the transcriptions. 
    }
    \centering
    \begin{tabular}{l cc cc}
        \toprule
        \multirow{2}{*}[-3pt]{\textbf{Method}} & \multicolumn{2}{c}{\textbf{COOLTEST}} & \multicolumn{2}{c}{\textbf{ASCEND}} \\
        \cmidrule(lr){2-3} \cmidrule(lr){4-5} 
        & \textbf{Man.} & \textbf{Eng.} & \textbf{Man.} & \textbf{Eng.} \\
        \midrule
        Whisper-Large-v2 & 13.23 & 20.64 & 24.68 & 47.84 \\
        \midrule
        KD--Full Data & 11.93 & 22.41 & 20.87 & 36.31 \\
        KD--Trivial Method & 11.74 & 21.77 & 17.61 & 33.88 \\
        \midrule
        K\(^2\)D--Direct \({\rm MER}\) & 11.31 & \textbf{17.93} & 18.56 & \textbf{33.12} \\
        K\(^2\)D--Direct \({\rm PER}\) & 10.77 & 18.57 & 16.62 & 33.35 \\
        K\(^2\)D--Composite \(\delta_{\rm comp}\) & \textbf{10.61} & 18.89 & \textbf{15.14} & 35.05 \\
        \bottomrule 
    \end{tabular}
    \label{tab:mer}
\end{table}

\subsection{Main Results}
\label{subsec:main_results}

We present the MER of each method on all the in-domain and out-of-domain testing sets in Table~\ref{tab:main}. 
As the table indicates, K\(^2\)D provides clear performance improvements on all the testing sets. 
Our method with the composite distance metric \(\delta_{\rm comp}\) (Eq.~\ref{eq:delta_composite}) produces improvements for over 17\% reduction rate on both of the in-domain testing sets. 
Furthermore, it demonstrates strong generalizability to different domains, especially to ASCEND, a publicly available code-switching realistic dataset. 
On the other hand, using the direct distance metrics, \({\rm MER}\) and \({\rm PER}\), give performance improvements as well. 
Note that the data usage of the two direct metrics is around \(60\%\), which is much less than the baseline methods. 
The result further strengthens the effectiveness of our method. 
In K\(^2\)D, we can perform simple and efficient data pre-filtering through cross-model validation, leveraging the knowledge from both the large and small models. 
Last but not least, our method achieves a five times faster generation speed than the teacher model.

\subsection{Performance Analysis on Each Language}
Next, we discuss the performance improvements our method brings to the two languages separately. 
We show the detailed evaluations of the CER for the Mandarin parts and the WER for the English parts on the code-switching datasets, COOLTEST and ASCEND, in Table~\ref{tab:mer}. 
We find out that using the direct MER as the distance metric produces the most improvements over English on both testing sets. 
In comparison, the composite metric yields the best performance in Mandarin. 
Applying K\(^2\)D can improve ASR performance for both languages across both the in-domain and out-of-domain testing sets. 

\begin{table}[!htb]
    \small
    \renewcommand{\arraystretch}{1.0}
    \caption{
        We present the detailed MER on the COOLTEST set and the repetitive hallucination counts detected using the \(n\)-gram method. Our results indicate that using the composite metric gives the fewest repetitive hallucinations. 
    }
    \centering
    \begin{tabular}{l cc cc}
        \toprule
        \multirow{2}{*}[-3pt]{\textbf{Method}} & \multirow{2}{*}[-3pt]{\begin{tabular}{c}\textbf{Rep.\(^\dagger\)}\\\textbf{Counts}\end{tabular}} & \multicolumn{3}{c}{\textbf{Detailed MER (\%)}} \\
        \cmidrule(lr){3-5}
        &  & \textbf{Del.} & \textbf{Ins.} & \textbf{Sub.} \\
        \midrule
        Whisper-Large-v2 & 110 & 4.53 &  4.07 & 5.35 \\
        \midrule
        KD--Full Data & 101 & 3.09 & 4.57 & 5.32 \\
        KD--Trivial Method & 47 & 2.75 & 4.53 & 5.29 \\
        \midrule
        K\(^2\)D--Direct \({\rm MER}\) & 40 & 3.20 & \textbf{3.69} & 5.07 \\
        K\(^2\)D--Direct \({\rm PER}\) & 45 & 2.80 & 3.74 & \textbf{5.00} \\
        K\(^2\)D--Composite \(\delta_{\rm comp}\) & \textbf{22} & \textbf{2.55} & 3.83 & 5.07 \\
        \bottomrule 
    \end{tabular}
    \label{tab:hallucinations}
\end{table}

\subsection{Investigation on Repetitive Hallucinations}
\label{subsec:investigation_on_repetitive_hallucinations}
As previously mentioned, repetitive hallucinations can be effectively detected using the \(n\)-gram method. 
In Table~\ref{tab:hallucinations}, we present the evaluation results and the counts of hallucinations identified by this method. 
We observe a significant reduction in repetitive hallucinations when data filtering is applied. 
Even with direct methods, the number of repetitive hallucinations drops substantially compared to using the entire dataset. 
The composite metric yields the fewest hallucinations, likely due to its combination of the \(n\)-gram and distance metrics, which leverages the strengths of both approaches during filtering. 
The deletion rate is also lowest with the composite metric, probably due to the reduced number of repetitive hallucinations.

\subsection{Validation Model Analysis}

\label{subsec:validation_model_analysis}
As we have mentioned in Section~\ref{subsec:models} We use Whisper-Base as the validator to provide additional \textit{knowledge} during cross-model validation. 
In this analysis, we wish to communicate that utilizing a smaller model for validation may provide efficiency, effectiveness, and robustness when the data is large-scale and realistic. 
First, we show that our method is efficient by comparing the time cost between using different variants of Whisper models as the \({\rm M}_{\rm validator}\). 
As shown in the table, Whisper-Base takes only 9 hours on 4 NVIDIA H100 GPUs, while Whisper-Medium takes over 30 times longer.

Next, we show that our proposed filtering method is effective in filtering out pseudo-labels with high error rates. 
We first define that a pseudo-label has a high error rate if its \(\text{MER} > 0.4\). 
We can then calculate the filter-out rate of the high-MER pseudo-labels for each cross-model validation method, which is composed of a \({\rm M}_{\rm validator}\) and a distance metric \(\delta\). 
The measurement of the filter-out rate is denoted as the \textbf{recall} in Table~\ref{tab:validation_model_analysis}. 
The \textbf{Max Recall} indicates the maximum filter-out rate of a valid \(\alpha\) that yields over 50\% of the data remaining in each cross-model validation method. 
Our results show that even with Whisper-Base, the max recall is similar to using the larger ones. 
This illustrates that we can achieve effective filtering on high-MER pseudo-labels under a reasonable filtering rate. 

Finally, we highlight that using Whisper-Base as the validation model is robust over the selection of \(\alpha\). 
By calculating the average recall (\textbf{Avg. Recall}) across different \( \alpha \in [0.1, 0.2, \ldots, 0.9]\), we find out that using Whisper-Base yields the highest average recall. 
This shows that the cross-model validation with Whisper-Base is less sensitive to the selection of \(\alpha\), enhancing the robustness of our method.

\begin{table}[!tb]
    \small
    \renewcommand{\arraystretch}{1.0}
    \caption{
        The performance comparison of using the different Whisper models as the validation model. 
        We approximate the time cost of Whisper-Small and -Medium by referencing the progress after one-hour generation. 
        The recall can be considered as the filter-out rate of the high-MER pseudo-labels on COOLTEST set. 
        \vspace{2pt}
    }
    \setlength\tabcolsep{4pt}
    \centering
    \begin{adjustbox}{width=0.95\columnwidth, center}
        
    \begin{tabular}{l c | cc cc}
        \toprule
        \multirow{2}{*}[-3pt]{\textbf{Validation Model}} & \multirow{2}{*}[-3pt]{\begin{tabular}{c}\textbf{Time}\\\textbf{Cost \(\downarrow\)}\end{tabular}} & \multicolumn{2}{c}{\textbf{Max Recall \(\uparrow\) }} & \multicolumn{2}{c}{\textbf{Avg. Recall \(\uparrow\)}}\\
        \cmidrule(lr){3-4} \cmidrule(lr){5-6}
        &  & \(\delta_{\rm MER}\) & \textbf{\(\delta_{\rm PER}\)} & \textbf{\(\delta_{\rm MER}\)} & \textbf{\(\delta_{\rm PER}\)} \\
        \midrule
        Whisper-Medium & 284 hr\hspace{2pt} & \textbf{0.97} & \textbf{0.97} & 0.63 & 0.62 \\
        Whisper-Small & 37 hr\hspace{2pt} & 0.91 & 0.85 & 0.60 & 0.54 \\
        \midrule
        Whisper-Base (Ours) \hfill & \textbf{9 hr}\hspace{2pt} & 0.91 & 0.86 & \textbf{0.69} & \textbf{0.66} \\
        \bottomrule 
    \end{tabular}
    \end{adjustbox}
    \label{tab:validation_model_analysis}
\end{table}




\section{Conclusion}
\label{sec:conclusion}

This work presents a novel framework, K\(^2\)D, for conducting practical and effective knowledge distillation with realistic data for code-switching ASR. 
Given the realistic data is unlabeled, noisy, and large-scale, performing data filtering might be essential. 
We introduce a novel method for efficiently filtering data based on the knowledge of a small auxiliary model. 
The process is based on the cross-model validation between the teacher model and the small model's transcriptions. 
Our results indicate the effectiveness of K\(^2\)D by surpassing the original teacher model with over 17\% and 30\% performance improvements on the code-switching in-domain and out-of-domain testing sets, respectively.
Furthermore, our method performs better than the baseline methods, which utilize the full dataset for training or conduct trivial \(n\)-gram-based data filtering. 
Last but not least, we conduct an analysis of our method, showing that the proposed filtering technique is efficient, effective, and robust. 
To our knowledge, we are the first ones to explore and enhance knowledge distillation in such a realistic scenario. 
We foresee that our method can facilitate more practical and effective knowledge distillation for ASR. 

\section{Acknowledgment}
This project has been powered by the computing resources from NVIDIA Taipei-1. 
Furthermore, we would like to express our gratitude to Hsuan-He Chang, Yi Chen, Yi-Hua Chen, Haw-Yang Foo, Wen-Yen Huang, Heng Hsu, Bo-Cheng Ke, Pei-Jhen Lan, Fang-Yu Liu, Ka-Sin Thoe, Yan-Tong Lim, Pin-Yu Lu, and Yu-Chi Peng for their efforts in data annotation for our in-house testing set of this project.

\bibliographystyle{IEEEbib}
\bibliography{strings,refs}

\end{document}